\def\that{{\widehat t}\,} 

\def\leaderfill{\leaders\hbox to 1em{\hss-\hss}\hfill}
 
\def\etal{{\it et al.}}

\def\msun {{ \rm \, M_\odot}}

\def\lsim{\mathrel{\mathpalette\@versim<}}
\def\gsim{\mathrel{\mathpalette\@versim>}}
\def\@versim#1#2{\lower0.2ex\vbox{\baselineskip\z@skip\lineskip\z@skip
  \lineskiplimit\z@\ialign{$\m@th#1\hfil##\hfil$\crcr#2\crcr\sim\crcr}}}

\def\et{{\frenchspacing\it et al.}\ }
\def\simlt{\hbox{ \rlap{\raise 0.425ex\hbox{$<$}}\lower 0.65ex\hbox{$\sim$} }}
\def\simgt{\hbox{ \rlap{\raise 0.425ex\hbox{$>$}}\lower 0.65ex\hbox{$\sim$} }}

\def\that{{\hat t}}

\def\etal{{\it et al.}}

\def\msun{M_\odot}
\def\vperp{v_{\perp}}

\documentstyle[11pt,paspconf]{article}

\begin{document}

\title{The Case for a Next Generation LMC Microlensing Survey}

\author{Christopher W. Stubbs\altaffilmark{1}}

\altaffiltext{1}{
Department of Astronomy and Department of Physics, 
University of Washington, Seattle WA USA 
}

\begin{abstract}
Microlensing surveys
search for the transient brightening of a background star 
that is the signature of 
gravitational lensing by a foreground compact object. This 
technique is an elegant way to search for astrophysical 
candidates that might comprise 
the dark matter halo of the Milky Way. While the current projects have 
successfully detected the phenomenon of microlensing and have reported
many important results, the relatively large event rate reported towards the 
LMC remains a puzzle. The first step in resolving this mystery is determining
the {\it location} of the excess lensing population. This will require 
a microlensing survey with an order of magnitude increase in sensitivity
over current projects. I summarize the present status of microlensing 
surveys, and present (and advocate!) a next-generation program that should
be capable of unambiguously determining whether the dark halo of the 
Galaxy is indeed
made up of MACHOs, or whether the observed events are due to previously
unappreciated ordinary stellar populations. 
\end{abstract}


\section{Introduction}

A convergence of progress in both computing and detector technology in 
the late 1980's led a number of groups to undertake
searches for gravitational microlensing. A primary motivation 
for much of this work was the elegant idea (Paczynski 1986) of searching for
Galactic dark matter in the form of MAssive Compact Halo Objects (MACHOs) 
by looking directly for their gravitational effects. This has the advantage 
of using the one thing we know about dark matter; it exerts a gravitational 
influence on its surroundings.  Comprehensive reviews of the technique and 
initial results have been presented by Paczynski (1996) and 
Roulet and Mollerach (1997).

Although the microlensing surveys are frequently 
characterized as
searches for Baryonic dark matter, we should bear in mind that these
projects are sensitive to any object that gravitates like a point 
mass. MACHOs need not necessarily be made of Baryonic matter.
 
The basic idea in a microlensing dark matter hunt is to monitor a 
population of background stars and search 
for the signature of gravitational lensing due to foreground compact objects. 
If the foreground object passes close to the line of sight to a background 
star, gravitational lensing produces multiple images of the source star. 
Due to the relatively short (Galactic-scale!) distances in the case of 
microlensing, the multiple images are not individually resolvable, 
as opposed to lensed quasars which can have image separations of order 
arc seconds. Since the source, lensing object and detector are in relative 
motion the observable effect is a transient brightening of the background 
star, with an apparent amplification $A(u)$ given by 
\begin{eqnarray}
\label{eq-amp}
  A(u) & = {u^2 + 2 \over u \sqrt{u^2 + 4} },  \\
    u & = b / R_E,   \\
  R_E & = \sqrt{4Gm D_1 D_2 \over c^2 \,(D_1 + D_2) }
\end{eqnarray}
where $b$ is the separation of the lens from the undeflected line of sight,
$R_E$ is the Einstein radius (in the lens plane),
$D_1$,$D_2$ are the distances from the lens to the source and 
detector, respectively,  
and $m$ is the lens mass.

The fraction of background stars 
that lie within one Einstein radius of a foreground lens is a 
dimensionless number: the optical depth $\tau$.  For an isothermal halo 
with a typical velocity $\beta = v/c \sim 10^{-3}$, we expect $\tau 
\sim 10^{-6}$. This means that a microlensing survey must monitor many
millions of stars in order to have a reasonable event detection rate. 
 
The duration $\that $ of a microlensing event is determined by the lens' 
position, its mass, and its transverse velocity. It is difficult to 
disentangle these for any single event.   Real-time detection and 
follow-up observations of exotic events are helping to lift this 
degeneracy in certain cases, however, as described below.
For a fixed total mass of 
lensing objects in the Galactic halo, a population of low mass lenses 
produces short events, while if the MACHOs were more massive there 
would be fewer, but longer, events.

Microlensing surveys are being carried out towards stars in the Galactic
center, and towards our near-neighbor galaxies the Large Magellanic Cloud
(LMC) and Small Magellanic Cloud (SMC). More speculative work is
being carried out towards M31 (Andromeda) and other targets.

Using wide-field CCD arrays and 1 meter class telescopes, three groups 
made essentially simultaneous announcements (Udalski \et al 1993, 
Alcock \etal 1993, Aubourg \etal 1993) of the 
detection of microlensing.    
At the time of this writing (Fall 1998) the total number of 
reported microlensing events exceeds 300. The overwhelming majority 
of these are seen towards the Galactic center, presumably due to 
lensing by ordinary objects in the disk or bulge of the Galaxy.  
The ratio of the number of microlensing events reported thus far towards the 
Galactic center to the number seen towards the LMC and 
SMC is very roughly given by $(N_{Center} : N_{LMC} : N_{SMC}) \sim (100 : 
10 : 1)$. 
The event rates towards the LMC and SMC probe directly the amount of lensing
material along those lines of sight, which may be dominated by dark matter. 
The microlensing rate towards the Galactic Center also has a bearing on the
Galactic dark matter, as it can be used to map out the mass distribution in 
the disk. This can help determine the extent to which the Galactic rotation
curve is supported by disk matter. 

\begin{figure}[!htb]
\plotfiddle{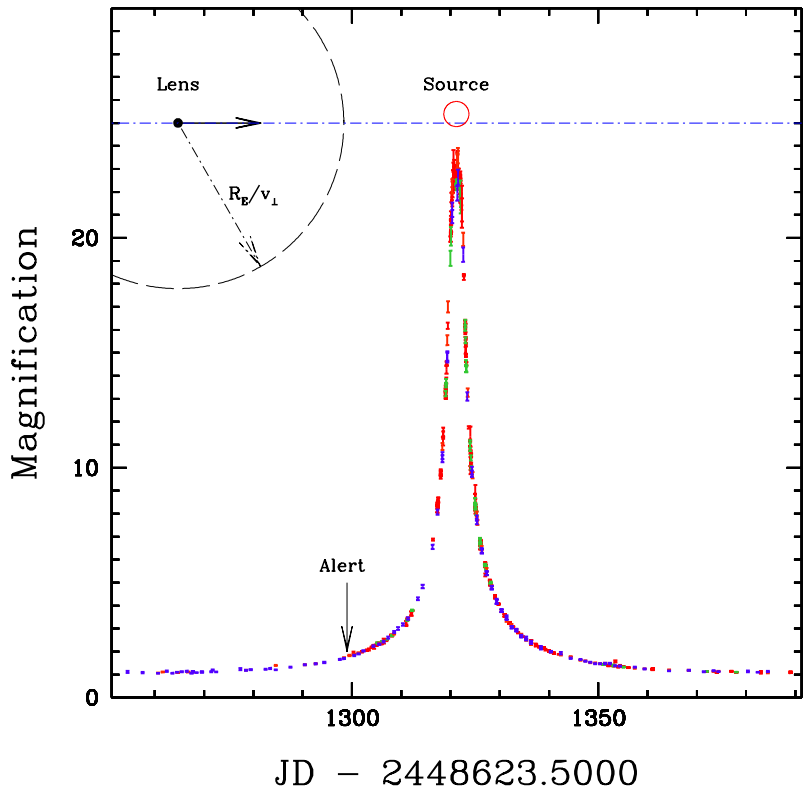}{105truemm}{0}{55}{55}{-180}{-80}
\caption{
Light Curve of MACHO event 95-30, an exotic ``Finite Source'' event.
The amplification as a function of time (in days) is shown. Also shown
to scale are the size of the source star and the Einstein radius of the 
lensing object, which traversed the face of the star. Intense monitoring
of exotic 
events can be used to lift the degeneracy between lens
mass, position, and transverse velocity.  
}
\end{figure}

\subsection {Lifting Degeneracies with Exotic Events}

The formulation for $A(u)$ given above assumes a point lens, 
a point source, and a detector that are in relative inertial motion. If any
one of these assumptions breaks down, the observed light curve can depart from
the form given by Eq 1. These perturbations, sometimes very dramatic, 
can convey additional information about the lens mass, location or 
velocity. One of the pleasantly surprising aspects of the 
microlensing field has been the speed with which the rich phenomenology
of exotic microlensing was observationally established.  
\vskip0.1in
Examples of exotic microlensing that have been seen include:
\vskip0.1in
1) Finite source effects, where the lens passes very close to (or
even across the face of) the background star. In this case the peak
amplification is attenuated since the source of light is distributed
across an extended region.  An example of such an event is shown in 
Figure 1, taken from Alcock \etal, 1997b. 
Finite source effects provide a way to 
determine $\theta = \phi_e / \phi_{source}$,  
the angular size of the Einstein radius, $R_{E}$,  in units of the 
source star's angular extent.  The projected transverse velocity of the 
lens can also be determined.     
\vskip0.1in

2) Parallax effects, due to the acceleration of the line of sight as the earth
orbits the sun. This can be used to establish a relationship between the 
lens' mass and its position along the line of sight.  
\vskip0.1in

3) Xallarap effects, the complement of parallax effects, where the 
source star is orbiting a companion object. (Xallarap is ``parallax"
backwards).   
\vskip0.1in

4) Binary lenses, where a compound system acts as a lens. In this case, 
dramatic structures can be seen in the light curve, as the caustic
lines of the asymmetrical lens pass across the source star. By
measuring the time it takes for a caustic line to pass across a 
star of known diameter, the projected velocity of the lens can be 
measured. An example of a binary lensing event is shown 
in Figure 2. 
\vskip0.1in

\begin{figure}[!htb]
\plotfiddle{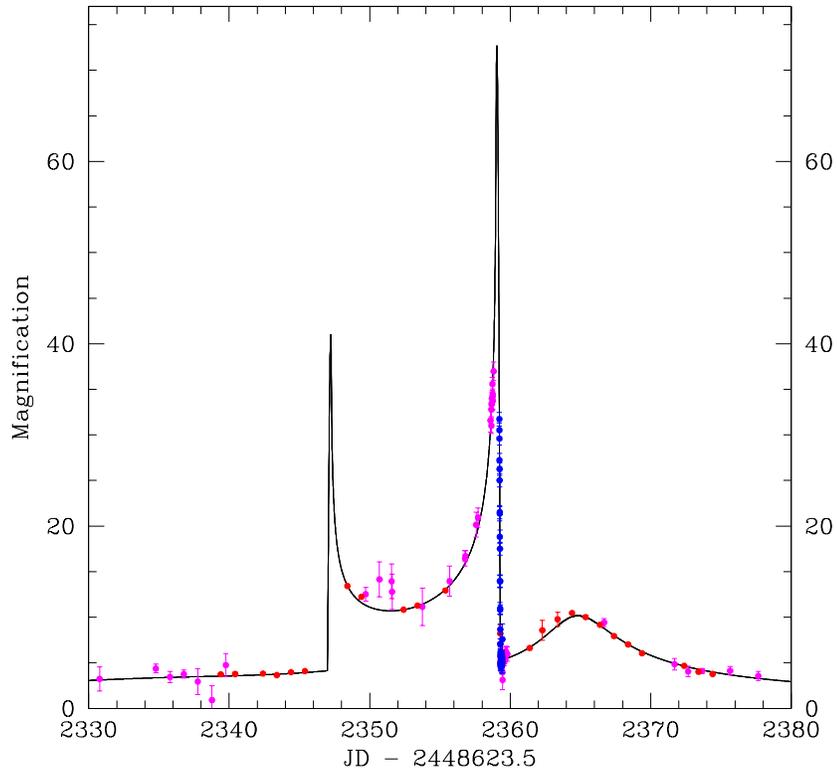}{80truemm}{0}{55}{55}{-170}{-160}
\vskip 1.0in
\caption{
Example light curve from a binary lens, in this case MACHO SMC-98-1. 
The dramatic changes in the light curve occur
when caustic lines cross in front of the source star, and the transit time
can be used to measure the angular transverse velocity of the lens, 
given the size of the source star. 
}
\end{figure}

Much of the data shown in Figure 2 were obtained after the real-time 
detection of the anomaly in the light curve (Alcock \etal, 1998b,
Afonso \etal, 1998). Currently, not only are
the majority of microlensing events detected in 
real time (meaning within a few hours of the image being taken), but
anomalies in the light curves of ongoing events
are also sought, giving rise to ``Level 2'' 
Alerts.  See {\tt http://www.darkstar.astro.washington.edu}
and {\tt http://www.astrouw.edu.pl/$\sim$ftp/ogle/ogle2/ews/ews.html}
for descriptions of ongoing events.    

The observed frequency and characteristics of these various exotic events  
will play an important role in the next-generation survey described below. 

\section {Current and Future Results from Existing Microlensing Surveys}

The extensive monitoring programs carried out by the microlensing 
surveys have generated a prodigious amount of raw image data. 
The MACHO project, for example, has generated over 6 Terabytes of CCD 
image data.   The frames are subjected to 
photometric analysis, and the resulting light curves are 
scanned for evidence of microlensing.  
Approximately one object per
thousand in the MACHO database exhibits significant variability. While
intrinsic stellar variability is a nuisance when searching for microlensing, 
this data set is an unprecedented resource for the study of variable 
stars. This paper, however, will focus on dark matter.   

The microlensing projects have collectively produced two results of 
significance regarding the Galactic dark matter.
In order of significance they are:  

\vskip0.1in

1. The lack of short events in the joint MACHO + EROS data set 
excludes any compact objects in the mass range 
$10^{-7} <  M/\msun < 10^{-3}$ from 
comprising a significant fraction of the Galactic dark 
matter (Alcock \etal, 1998).
\vskip0.1in

2. The MACHO team has reported (Alcock \etal 1997a)
an excess of events towards the LMC, 
relative to the number expected from known stellar populations. 
Cast in terms of an optical depth, 
$\tau^{200}_{2}=2.9^{+1.4}_{-0.9} \times 10^{-7} $ vs. the 
$ \sim 0.5 \times 10^{-7}$ expected from microlensing due to known
stellar populations.
\vskip.1in

\begin{figure}[!htb]
\plotfiddle{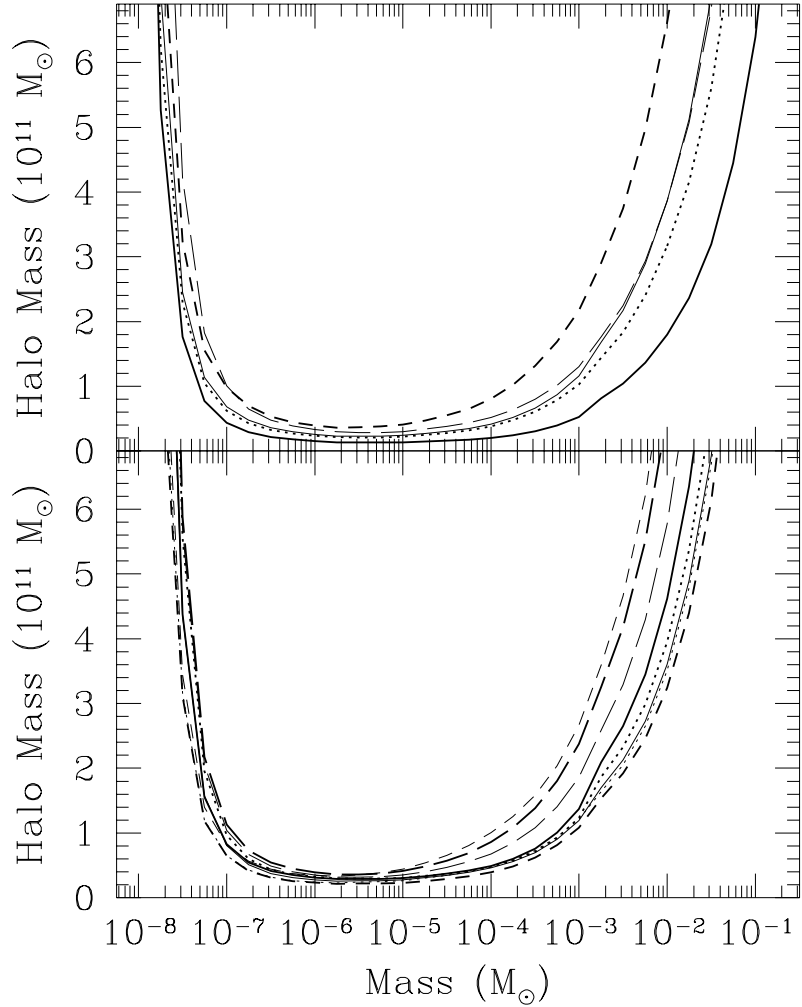}{135truemm}{0}{70}{70}{-250}{-110}
\caption{
Exclusion plot from the joint MACHO + EROS search for
short microlensing events.  The curves in the two
panels show constraints
for a variety of halo models. The regions above the lines are
excluded at greater than 95\% confidence. Note for comparison that a mass
(within 50 kpc) of $\sim 4.5 \times 10^{11}$ is 
typically predicted for models of the
Galactic halo.
}
\end{figure}

\subsection{The Null Result- No Short Events}

The exclusion plot from the joint EROS + MACHO search (Alcock \etal 1998) 
for 
events with durations less than 20 days is shown in Figure 3. 
This null result is very 
important progress in the ongoing search for dark matter, as it 
excludes much of the previously ``favored'' astrophysical mass regime, 
particularly the notoriously elusive brown dwarf population below the 
0.08 $\msun$ nuclear ignition threshold.  

This exclusion plot, which spans over five decades in mass at interesting 
confidence levels, is a triumph of the microlensing surveys. 
On the basis of this null result alone, the microlensing projects have 
been a great success! 

\subsection {The LMC Events: The Detection of Dark Matter, Or Is It Just Stars?}

The MACHO project has published (Alcock \etal, 1997a) a 
determination of the optical 
depth $\tau^{200}_{2}$ towards the LMC, 
for events with durations between 2 and 200 days. 
Interestingly, the observed optical depth and the event rate
towards the LMC exceed the values
expected from known stellar populations. In this context ``known''
refers to populations that were used for optical depth predictions  
made before the results were presented. 

Since the distribution of durations for halo microlensing events depends 
on the velocity, mass and spatial distributions of the lensing population, 
one must in general make assumptions about 
two of these in order to constrain the 
third. If we assume an isothermal halo population of lensing objects, 
and a rotation curve that is flat at the solar circle, taken at face value
the MACHO result corresponds to having detected roughly half of the 
halo dark matter, in objects with masses of about half a solar mass.   

The alternative to the dark matter interpretation, of course, is lensing
by some stellar population. There are a number of possibilities for 
producing a rate of lensing events towards the LMC that exceeds the 
predictions of double-exponential axisymmetric models of the Galaxy. 
Any one or some combination of them might well be responsible for the 
observed lensing events. The proposed populations include:
\vskip0.1in
\begin{description}

\item[1)] Stars in the LMC,   
\vskip0.1in

\item[2)]
An intervening dwarf galaxy,   
\vskip0.1in

\item[3)]
The spheroidal component of our Galaxy,  
\vskip0.1in

\item[4)]
A wisp of tidal debris from the interaction between the LMC and 
the Galaxy,   
\vskip0.1in

\item[5)]
A severe warp in the disk of the Galaxy, and  
\vskip0.1in

\item[6)] A thick disk component of our Galaxy. 
 
\vskip0.1in
\end{description}

All dark matter searches face the challenge 
of discriminating between background
processes and the signal from dark matter. Microlensing is no exception. The
survey teams have successfully learned how to discriminate microlensing 
events from intrinsic stellar variability, and the challenge now is to 
distinguish between microlensing due to dark matter objects and microlensing
due to stars.  

In my opinion there are two logically distinct stepping stones on the 
path forward.  First, we need to 
determine with high confidence whether the microlensing events are due to 
Galactic dark matter. If so, and only then,  
we will need to establish how much of the halo is in the form of MACHOs, and 
ascertain the nature of these lensing objects (white dwarfs, 
primordial black holes, MACHinos...).  

\section{Distinguishing Between Lensing Populations}
 
The main point of this paper is that the various proposed 
lensing populations have observable differences which,  
given enough events, will allow us to determine 
what is responsible for the LMC events.  
Most importantly, we should be able to discriminate halo MACHOs from
the various stellar alternatives, using a variety of distinguishing
characteristics. 

\subsection{Luminosity}

If ordinary stars are responsible for the microlensing events, then
these stars should be visible in careful studies of color-magnitude diagrams, 
proper motions, and star counts. Some of this work is already
under way, and is an important complement to the microlensing surveys. 

\subsection{Spatial Distribution}

The spatial distributions of the proposed stellar lensing populations
differ from the $\rho_{DM}(r) \sim 1/r^2$ scaling
expected in typical dark matter halo
models. Both along the line of sight and across the sky there will
be observable differences between the various alternatives. Over the 
$\sim$ 120 square degrees of the LMC we would expect a fairly uniform
(after correcting for detection efficiency) distribution of halo lensing
events, whereas tidal tails and foreground stellar associations would
likely show spatial structure on this scale. 
This of course assumes there is no substructure
in the phase space distribution of the dark matter. 

The location of the 
lenses along the line of sight is also measurably different for the 
various populations.  For example, a nearby population of thick disk lenses
will show more parallax effects than would a more remote class of objects. 
Conversely, lenses in the LMC would exhibit more sensitivity to the
xallarap effect. High accuracy followup photometry would help determine the 
relative incidence of these light curve fine structure effects.    

\subsection{LMC stellar density}

The various stellar populations will also have lensing 
rates that have different  
scalings with LMC stellar density. For example, the case of 
LMC-LMC lensing should (after correcting for event detection efficiency) 
have a very different dependence on LMC stellar density, $ \rho_{\star}$, 
than lensing by
foreground populations. Roughly, we would expect LMC-LMC lensing to show
an event rate that scales as $\epsilon \rho_{\star}^{2}$, where $\epsilon$ 
is a density-dependent detection efficiency factor. Events from halo MACHOs, 
on the other hand, would likely scale as $\epsilon  \rho_{\star}$, a linear
rather than quadratic dependence on background stellar density. 

\subsection{Kinematics}

Finally, there are big differences in the projected velocity of the lens, 
as seen in the plane of the source. 
The relevant observable quantity is $v_{proj}$, the projected transverse
velocity with which the lens sweeps across the source star, as seen in the
frame of the source. For a lens at a dimensionless distance 
$x= {D_2 \over (D_1 + D_2)}$ 
along the line of sight,  
$v_{proj} = \vperp /x$, where $\vperp$ is the lens' velocity
component transverse to the line of sight. Nearby lenses have 
large projected velocities, 
compared to a more distant lens moving at the same transverse velocity. 

This is a powerful discrimination tool, as shown in the recent case of 
event MACHO 98-SMC-1, where the location of the binary lens was inferred
from a measurement of the caustic crossing time. 

\subsection{How to Tell Where the Lenses Are}

Table 1 summarizes the distinguishing characteristics of the different 
populations. Given enough events, particularly ones detected in real
time with adequate followup photometry, we should be able to discriminate
between the alternative lensing populations. 

\begin{table}
\caption{Distinguishing Characteristics for Various Lensing Populations}
\begin{center}
\begin{tabular}{lcccc}
\tableline\tableline
 & Thick Disk & Foreground Stars & LMC Stars & Dark Matter \\
\tableline
Parallax & few \% & Minimal & No & $\sim$ 1 \%  \\
Xallarap & No     & Minimal & Yes & Minimal \\
$\epsilon \rho_{\star}$ Scaling & Linear & Linear & Quadratic & Linear \\
Spatial  & Uniform & Non-Uniform & Non-Uniform & Uniform \\
$v_{proj}$(km/sec) & 5000 & 300/$x$ & 80 & 1000 \\
Luminous? & Yes & Yes & Yes & No \\
\tableline\tableline
\end{tabular}
\end{center}
\end{table}

How many events will be needed? We certainly need to be out of the 
present regime where the Poisson statistics of a few events preclude
our investigating dependences on stellar-density and spatial 
distributions.  Also, roughly 10\% of the Galactic center events seen to date 
exhibit some sort of exotic structure in their light curves, and we
hope to exploit this towards the LMC as well. Both of these considerations
argue in favor of event totals in the hundreds, an order of magnitude
increase from the present sample. Will the present surveys provide
this? I think not.  
 
\subsection{A Look Ahead: Microlensing in Y2K} 

Current plans call for the MACHO project to terminate operation
at the end of 1999, and the EROS team also intends to shut down in a similar
time frame. 
By the year 2000, the existing microlensing 
projects will likely have
detected tens of LMC events, a handful towards the SMC, and 
many hundreds of events towards the Galactic center. We should have
by then contour maps of $\tau(\ell, b)$, the optical depth towards 
the Galactic center as a
function of Galactic latitude and longitude. These will provide important
new information on the structure of the Galaxy, and will impact our
interpretation of the Galactic rotation curve. In addition, the lensing
events of Galactic bulge and disk stars will provide unprecedented 
opportunities for high resolution spectroscopy and (by searching for
perturbations of the light curves) opportunities to search for 
relatively low mass planetary companions to the lensing stars. 

The optical depth towards the LMC will likely have been measured at 
the 20\% level, and that towards the SMC at something like the
100 \% level. The nature of the lensing populations will, in my
opinion, remain a mystery.   

This is because, based on existing detection rates, the current projects will 
simply not have generated enough events by the 
year 2000 to meet the criteria given above. A new project, with 
at least an order of magnitude increase in sensitivity, is needed, 
and can be carried with existing technology (Stubbs 1997). 
Rather than achieving the required event total by operating the existing 
surveys for another decade, a next-generation project would detect a 
comparable number in the first year of operation. 

\section{A Next Generation Microlensing Survey}

An appropriate figure of merit for a microlensing survey is simply the
number of stars that can be monitored in a given amount of time.  
This is obtained from the expression that gives the Signal-to-Noise ratio
for the measurement of the flux $\Phi_{star}$ from a single star, 
$SNR ={{\Phi_{star}D^2 (QE) t} \over {\sqrt{\sigma^2 \Phi_{sky} D^2 (QE) t}}}$, 
where $D^2$ is the telescope effective aperture, $QE$ is the detector
efficiency, $t$ is the measurement time, $\Phi_{sky}$ is the sky background flux
, and $\sigma$ is the seeing. (The approximation of objects fainter than
sky is appropriate, as shown below). This expression can be rearranged
to determine the time needed to obtain a given Signal-to-Noise ratio, 
so that the overall figure of merit, including now a term FOV that 
gives the field of view of the camera system, for a microlensing survey
is $FOM = {{D^2 (QE) FOV} \over {\sigma^2 \Phi_{sky}}}$.

No big surprise here. A survey benefits from a dark site with good 
seeing, and a high throughput wide field camera. This expression does
allow us to make a quantitative comparison between existing and 
potential future projects, however.

\subsection {A Concrete Proposal}
The MACHO project uses a pair of 4K x 4K pixel front-illuminated
cameras on a 1.3 meter
telescope on the outskirts of Australia's capital city. 
By comparison 
the proposed new survey will be carried out at one of the 
world's premier astronomical sites, on a much larger aperture telescope than
any used by an existing survey, 
with a state-of-the-art wide field camera system.  
We are working to establish a next-generation project using the
DuPont 2.5 meter telescope at Las Campanas,
equipped with a state-of-the-art mosaic of 2K x 4K thin CCDs. 
Table 2 lists a comparison of the relevant specifications of the two systems, 
for the elements that enter into the Figure of Merit given above. 

\begin{table}
\begin{center}
\caption{Figure of Merit Comparison}
\begin{tabular}{lccc}
\tableline\tableline
 & MACHO & Next Gen Survey & FOM Gain \\
\tableline
Seeing & 2 \arcsec & $<$1 \arcsec & $>$4 \\
CCD QE & thick & thin & 2  \\
Aperture & 1.3m & 2.5m & 3.5 \\
Field & 0.5 sq deg & 1 sq deg & 2  \\
Sky & & & $>$2 \\
\tableline
Nominal Figure of Merit Gain & & & $>$100  \\
\tableline\tableline
\end{tabular}
\end{center}
\end{table}

While it is often overly optimistic to count on a performance improvement
given by a simple figure of merit, it is clear from Table 2 that an order
of magnitude increase in system throughput is readily achievable. This implies
that the next-generation survey described here should be able to monitor
at least ten times as many stars as the MACHO project does, in an equal amount
of time. Reality is slightly more complicated. There are a finite number of
stars in the LMC bright enough to usefully monitor with the system 
described here, and their luminosity distribution will determine whether
there are in fact ten times as many detectable events to be seen. 

Gould (1998) has calculated these effects in some 
detail, and comes to the 
reassuring conclusion that in fact an order of magnitude increase in event tally
is achievable with the system parameters summarized in Table 2. Attaining
this will require monitoring all regions in the LMC with a surface
brightness that exceeds 24.5 mag/arcsec$^2$. 
 
A tight coupling between the survey's detections
and followup telescopes will be an 
important ingredient in extracting the maximum information from exotic
microlensing events. A network of 1m class followup telescopes would
complement the next-generation survey system.
Selected events would ideally also be monitored with 
adaptive or space-based imaging systems, to achieve high angular resolution
and thereby minimize the effects of blended unlensed light. 

\section {Conclusion}

The microlensing technique is the only dark matter search that 
presently shows a persistent signal. 
It is imperative that we determine whether the 
observed event excess is due to a halo population of lenses. This is 
achievable,  with an appropriate commitment of existing telescope 
resources augmented with state-of-the-art CCDs and appropriate 
computing power.   

If the outcome of this next-generation project shows that ordinary stars
are responsible for the lensing excess, this will be strong
evidence in favor of particle physics dark matter, simply because
the astrophysical mass regime will have been effectively eliminated. 

On the other hand, if the lensing events are shown to be due to a 
halo population we will be confronting a new era in dark matter research.
The Galactic dark matter in the form of MACHOs will be a main focus
of our attention. 

While we can engage in an extended debate about the source of the 
events seen to date, this issue will not likely be resolved with 
existing surveys. This is an experimental question and we should
conduct the obvious next experiment. 

\acknowledgments
{
My colleagues on the MACHO team have made
crucial contributions to my understanding of the topics discussed
in this paper. I gratefully acknowledge many enlightening 
conversations with Andrew Gould, Charles Alcock, as well as 
Gus Oemler and his colleagues
at the Carnegie Observatories. Thanks also to Andrew Becker for his
comments on the manuscript and for his help with the figures. 
Generous support from the Packard Foundation is
gratefully acknowledged. This paper is dedicated
to my friend and colleague, Alex Rodgers.  
}


\end{document}